%% file: gravi_main.tex
\documentclass[a4paper,11pt]{article}
\pdfoutput=1 

\usepackage{jheppub} 
\usepackage{amsmath,amssymb,amsfonts}
\usepackage{graphicx}
\usepackage{longtable}
\usepackage{listings}
\usepackage{amsmath}
\usepackage{slashed}
\usepackage{ltxtable} 
\usepackage{booktabs}
\usepackage{subcaption}

\graphicspath{{Plots/}}

\newcommand{\GOSAM}{{\textsc{Go\-Sam}}}
\newcommand{\SAMURAI}{{\textsc{Sa\-mu\-rai}}}
\newcommand{\GOLEMVC}{{\texttt{Go\-lem95C}}}
\newcommand{\FEYNRULES}{{\texttt{Feyn\-Rules}}}
\newcommand{\UFO}{{\texttt{UFO}}}

\newcommand{\bea}{\begin{eqnarray*}}
\newcommand{\eea}{\end{eqnarray*}\noindent}
\newcommand{\be}{\begin{equation}}
\newcommand{\ee}{\end{equation}\noindent}
\newcommand{\bcen}{\begin{center}}
\newcommand{\ecen}{\end{center}}

\newcommand{\Gev}{\mbox{ GeV}}
\newcommand{\Tev}{\mbox{ TeV}}

\title{NLO QCD corrections to diphoton plus jet production through graviton exchange}

\author{Nicolas Greiner$^a$,
Gudrun Heinrich$^a$,
Joscha Reichel$^a$,
Johann Felix von Soden-Fraunhofen$^a$ 
}
\emailAdd{greiner@mpp.mpg.de}

\emailAdd{gudrun@mpp.mpg.de}

\emailAdd{joscha@mpp.mpg.de}

\emailAdd{jfsoden@mpp.mpg.de}

\affiliation[a]{Max Planck Institut f\"ur Physik, F\"ohringer Ring 6, 80805 M\"unchen, Germany}

\preprint{MPP-2013-226}

\abstract{
We present the NLO QCD corrections to the production of 
a photon pair in association with one jet, 
where the photons are stemming from graviton decay, 
within models of large extra dimensions.
Our results for the loop amplitudes are produced with the program 
\GOSAM{} for automated one-loop calculations. 
We show distributions for several observables for 4, 5 and 6 extra dimensions
and demonstrate that the differential K-factors are far from being uniform.
}

\begin{document}
\maketitle
\flushbottom

\section{Introduction}
\label{sec:intro}
\input intro

\section{Calculational framework}
\label{sec:calculation}
\input calculation

\section{Phenomenological results}
\label{sec:results}
\input results

\newpage
\clearpage

\section{Conclusions}
\label{sec:conclusion}
\input conclusion

\acknowledgments
We would like to thank all the members of the GoSam collaboration
for contributions to code development 
and constructive comments.  
We also thank Dorival Goncalves-Netto and Dieter Zeppenfeld for interesting  discussions.
We further acknowledge computing support of the Rechenzentrum Garching.





\input gravi_main.bbl

\end{document}

%% file: intro.tex
The discovery of a  new boson at the LHC~\cite{Aad:2012tfa,Chatrchyan:2012ufa}
constitutes a big step towards unveiling the mechanism of electroweak symmetry 
breaking. 
However, it is still a major task  to make sure that this new particle is indeed  the 
Standard Model Higgs boson. Therefore it is important to scrutinize its 
coupling strengths, decay modes, and parity/spin quantum numbers.
For the latter, the currently analyzed data strongly point to a CP even spin zero 
particle~\cite{Aad:2013xqa,ATLAS-CONF-2013-40,ATLAS-CONF-2013-31,ATLAS-CONF-2013-29,CMS-PAS-HIG-13-005,CMS-PAS-HIG-13-003}.
While a graviton-like spin-2 resonance with mass around $125$\,GeV
is excluded at the 99\% confidence level~\cite{Aad:2013xqa,Ellis:2012jv,Ellis:2012mj}, 
models involving gravitons with masses in the TeV range are still a viable extension of the Standard Model. 
Even though graviton masses up to about 4 TeV are  excluded at 95\% CL for $\delta\leq 4$ extra dimensions 
with simple model assumptions~\cite{ATLAS:2011ab,Aad:2012cy,Chatrchyan:2011jx,Chatrchyan:2011fq}, 
this can definitely not be interpreted in the sense that the existence of gravitons in the context of 
models involving extra dimensions is ruled out.

In models with flat extra dimensions, in particular the ones proposed by 
ADD~\cite{ArkaniHamed:1998rs} with $D=4+\delta$ dimensions, only the gravitons  
propagate in the full $D$-dimensional space-time, while the Standard Model fields 
are confined to four dimensions.
If the $\delta$-dimensional space is assumed to be compactified on a torus with common radius $R$, 
the 4-dimensional  Planck scale $M_{\rm{Pl}}\sim 10^{19}$\Gev{} is an effective scale, related 
to the fundamental scale $M_D$ of quantum gravity by  \cite{ArkaniHamed:1998rs}
\be 
	M_{\rm{Pl}}^2 = 8\pi  R^\delta M_D^{\delta+2}.
	\label{eq:mp_eff}
\ee
For large values of the compactification radius $R$ it is therefore possible to have a 
fundamental scale $M_D\sim 1$\Tev. 
Alternatively, the string scale $M_S$ is often used instead of $M_D$
in the literature \cite{Han:1998sg}. 
The two scales are related by \cite{Gleisberg:2003ue}
\begin{align}
	M_S^{\delta+2} = (4\pi)^{\frac{\delta+2}{2}}    \Gamma ( \delta / 2)  M_D^{\delta+2}\;.
	\label{eq:msformula}
\end{align}
As the extra-dimensions are assumed to be compact, the discrete Fourier decomposition of the graviton states 
leads to a tower of Kaluza-Klein (KK) modes in four dimensions with spacings of the order of the inverse size of the extra dimensions. 
Within the ADD model\,\cite{ArkaniHamed:1998rs} of large extra dimensions, the number of expected KK graviton states 
is very large, thus compensating the smallness of the individual graviton couplings to Standard Model matter.  
The excited graviton states preferably decay into gauge bosons, rather than fermions, because the spin-2 nature implies that fermions cannot 
be produced in an s-wave\,\cite{Han:1998sg}.
Therefore, and due to the clean experimental signature, the decay of graviton modes into two photons is of particular phenomenological interest.

At the LHC, searches for extra dimensions in diphoton events have been carried out~\cite{ATLAS:2011ab,Aad:2012cy,Chatrchyan:2011jx,Chatrchyan:2011fq},
and have lead to lower limits on the scale $M_S$ between 2.5\Tev{} and 3.92\Tev, depending on the number of extra dimensions
and assumptions on model parameters. 
On the theory side, NLO corrections for the process $pp\to G\to\gamma\gamma$ have been calculated in \cite{Kumar:2008pk,Kumar:2009nn}, 
supplemented by a parton shower in \cite{Frederix:2012dp,Artoisenet:2013jma}, based on extensions to the {\tt MadGraph/MadEvent} framework 
for spin-2 particles\,\cite{deAquino:2011ix}. Ref.~\cite{Artoisenet:2013jma} also discusses the case where the coupling strength
of the graviton to matter fields and gauge fields is non-universal, while  Ref.~\cite{Frederix:2013lga} considers graviton decays into 
massive vector boson pairs rather than photons.
The experimental analyses searching for extra dimensions in diphoton events 
have been carried out using a constant K-factor based on the results of \cite{Kumar:2008pk,Kumar:2009nn} to account for NLO corrections.

Diphotons at the LHC will often be accompanied by one or more high-$p_T$ jets. 
Compared to the diphoton inclusive case, observables involving an extra jet offer better control on backgrounds 
and more information on the interaction dynamics. 
Therefore  precise predictions for the production of diphotons through the exchange of a 
spin-2 particle in association with a hard jet can help to derive improved 
limits on models involving graviton exchange. 
This is  particularly important  since the 
K-factors turn out not to be 
uniform over the range of the diphoton invariant mass distribution, which in general is used to derive exclusion limits.
Further, observables which serve to determine 
spin/CP properties of the object(s) leading to photons in the final state can be altered at NLO as new helicity 
configuration channels may open up, thus invalidating leading order studies.
The NLO QCD corrections to the production of stable gravitons in association with a jet have been calculated in 
\cite{Karg:2009xk}. 
The phenomenology of spin-2 resonances produced by vector boson fusion is studied at NLO QCD in ~\cite{Frank:2012wh,JessicaFrankDiplom}, 
in an effective Lagrangian approach.
NLO corrections to the Standard Model process $pp\to \gamma\gamma +$jet+X have been calculated in \cite{DelDuca:2003uz,Gehrmann:2013aga}, 
where the K-factors turned out to be rather large.

This article is organized as follows. In Section \ref{sec:calculation}, we discuss details of the calculation, in particular 
the treatment of the graviton propagator and the implementation of a general framework for spin-2 particles into 
the  one-loop program package \GOSAM{}.
In Section \ref{sec:results} we present  numerical results, followed by  our conclusions in Section \ref{sec:conclusion}.

%% file: calculation.tex
\subsection{The graviton propagator in large extra dimensions}

For the Feynman rules, we follow the conventions of \cite{Han:1998sg}. 
The  Lagrangian for the interactions of the  graviton Kaluza-Klein (KK) modes with the Standard Model matter 
proceeds via the energy-momentum tensor $T^{\mu \nu}$
\begin{align}
	{\cal L}_{Int} = - \frac{\kappa}2 \sum_{\vec n} G^{(\vec n)}_{\mu\nu} T^{\mu \nu}\;,
\end{align}
where $\kappa$ is related to the reduced Planck mass in 4 dimensions by $\kappa=\sqrt{16\pi}/M_{\rm{Pl}}$, 
and the graviton modes follow the equation
\begin{align}
	\left(\Box + \frac{4 \pi^2 {\vec n}^2}{R^2}\right) G_{\mu\nu}^{(\vec n)} = - \frac{\kappa}{2} T_{\mu\nu}\;.
\end{align}
%
The KK mode propagator can be split into two parts
\begin{align}
	i \Delta_{\mu\nu,\rho\sigma}(k,m_{\vec n}) =  \underbrace{\frac{i}{k^2-m_{\vec n}^2 + i\epsilon}}_{D(k^2,m_{\vec n})} B_{\mu\nu,\rho\sigma}(k,m_{\vec n})\;,
\end{align}
where $B_{\mu\nu,\rho\sigma}$ carries the Lorentz structure
\begin{align}
	B_{\mu\nu,\rho\sigma}(k,m) =&
	             \left(\eta_{\mu \rho} - \frac{k_\mu k_\rho}{m^2}\right) 
		     \left(\eta_{\nu \sigma} - \frac{k_\nu k_\sigma}{m^2}\right)
           + \left(\eta_{\mu \sigma} - \frac{k_\mu k_\sigma}{m^2} \right)
		     \left(\eta_{\nu \rho} - \frac{k_\nu k_\rho}{m^2}\right) \notag \\
         &  - \frac23  
	   \left(\eta_{\mu \nu} - \frac{k_\mu k_\nu}{m^2}\right)
	   \left(\eta_{\rho \sigma} - \frac{k_\rho k_\sigma}{m^2}\right)\;.
\end{align}
If all particles attached to the propagator are on-shell, 
the mass dependent terms in $B_{\mu\nu,\rho\sigma}(k,m)$ drop out.
For the calculation presented here, the on-shell condition is not always fulfilled, 
but we checked that the impact of the mass dependent terms is numerically 
negligible\footnote{The same  has been found in \cite{Gleisberg:2003ue}.} 
and therefore did not include them in our calculation.
In this case the summation over the graviton states  in $D(s,m_{\vec n})$, leading to
\be
D(s) = \sum_{\vec n} \frac{i}{s-m_{\vec n}^2 + i \epsilon}\;,
\ee
can be performed independently from the  $B_{\mu\nu,\rho\sigma}$ part carrying the Lorentz structure.
Further, we use the assumption that the widths of the KK modes are negligible, 
as the dominant effects come from the almost on-shell production of KK modes.
The discrete spectrum of the KK  modes can be approximated by an
integral over a mass density, as the  KK modes are very contiguous\,\cite{Han:1998sg,Giudice:1998ck}. 
The density as a function of the mass $m_{\vec n}$ is given by
\be
\rho(m_{\vec n})=\frac{R^{\delta}m_{\vec n}^{\delta - 2}}{(4\pi)^{\delta/2}\Gamma(\delta/2)}\;,
\ee
leading to \,\cite{Han:1998sg}
\begin{align}
	D(s) \to  \int_0^{M_S} d\,m_{\vec n}^2\,\rho(m_{\vec n})\,\frac{i}{s-m_{\vec n}^2+ i \epsilon}
	= \begin{cases} \frac{ s^{\delta/2-1}}{2M_{s}^{\delta + 2} G_N } \left( \pi + 2 i \, I(\frac{M_S}{\sqrt{s}}) \right) & \text{for\ } s>0 \\
		\frac{ (-s)^{\delta/2-1}}{2M_{s}^{\delta + 2} G_N} (-2i)\, I_E(\frac{M_S}{\sqrt{-s}}) & \text{for\ } s<0 
	\end{cases}
	\label{eq:propagator}
\end{align}
with 
\begin{align}
	I(x) = \begin{cases} - \sum_{k=1}^{\delta/2-1} \frac1{2k} x^{2k} - \frac12 \log(x^2-1)& \text{if\ } \delta \text{ even} \\
		- \sum_{k=1}^{(\delta-1)/2 } \frac{1}{2k-1} x^{2k-1} + \frac12 \log\left( \frac{x+1}{x-1} \right) & \text{if\ } \delta \text{ odd}
	\end{cases}
\end{align}
and
\begin{align}
	I_E(x) = \begin{cases} (-1)^{\delta/2 + 1} \left( \sum_{k=1}^{\delta/2-1} \frac{(-1)^k}{2k} x^{2k} + \frac12 \log(x^2+1)  \right) 
		& \text{if } \delta \text{ even} \\
		(-1)^{(\delta-1)/2} \left(   \sum_{k=1}^{(\delta-1)/2 } \frac{(-1)^k}{2k-1} x^{2k-1} + \frac12 \tan^{-1}(x) \right)   & \text{if } \delta \text{ odd} \,.
	\end{cases}
\end{align}
The UV cutoff $M_S$ is introduced as the effective theory approach loses its validity beyond the scale $M_S$.

\subsection{Details of the implementation}

The  virtual corrections are calculated by 
the one-loop generator \GOSAM\,\cite{Cullen:2011ac}.
The program combines cut-based integrand reduction 
techniques~\cite{Ossola:2006us,Ellis:2007br,Mastrolia:2008jb,Mastrolia:2010nb,Heinrich:2010ax} 
with improved tensor reduction methods\,\cite{Binoth:2008uq,Cullen:2011kv}.
The rational part can be calculated algebraically within {\sc GoSam} in an automated way.
\GOSAM\, is publicly available at {\tt http://gosam.hepforge.org/}.

In more detail, the code generation proceeds as follows. \GOSAM{} reads an input card 
edited by the user and generates the diagrams 
and the corresponding expressions for the loop amplitudes,
using {\tt QGRAF}\,\cite{Nogueira:1991ex} and {\tt FORM} \,\cite{Vermaseren:2000nd,Kuipers:2012rf}
in combination  with {\tt Spinney}\,\cite{Cullen:2010jv} for the spinor algebra, 
and {\tt haggies}\,\cite{Reiter:2009ts} for optimisation and automated code generation.
The reduction can be performed in several ways, using integrand reduction based on 
\SAMURAI{}\,\cite{Mastrolia:2010nb} or tensor reduction based on 
\GOLEMVC{}\,\cite{Binoth:2008uq,Cullen:2011kv}, 
or a combination of the two\,\cite{Heinrich:2010ax}.
The basis integrals are taken from \GOLEMVC{} or {\tt OneLOop}\,\cite{vanHameren:2010cp}.

For models beyond the Standard Model, model files generated by \FEYNRULES{}\cite{Christensen:2008py} 
in the \UFO \,(Universal Feynrules Output)\,\cite{Degrande:2011ua} format 
can be imported directly by \GOSAM.
For the process under consideration, \GOSAM{} has been extended to be able to deal with 
spin-2 particles, 
and a version of the \GOLEMVC{} library is used which contains 
integrals where the tensor rank $r$
exceeds the number $N$ of propagators\,\cite{golemhighrank}. 
The  extension of the reduction at integrand level to cases where 
$r=N+1$ has been worked out in \cite{Mastrolia:2012bu}.

For the  QCD part, we work in the Feynman gauge, and renormalisation has been done in the $\overline{MS}$ scheme.
As the interaction between the Kaluza-Klein modes of a graviton and the Standard Model particles 
is via the energy-momentum tensor which is a conserved quantity, no further renormalisation procedure is needed.

\begin{figure}[htb]
\centering
\includegraphics[width=0.7\textwidth]{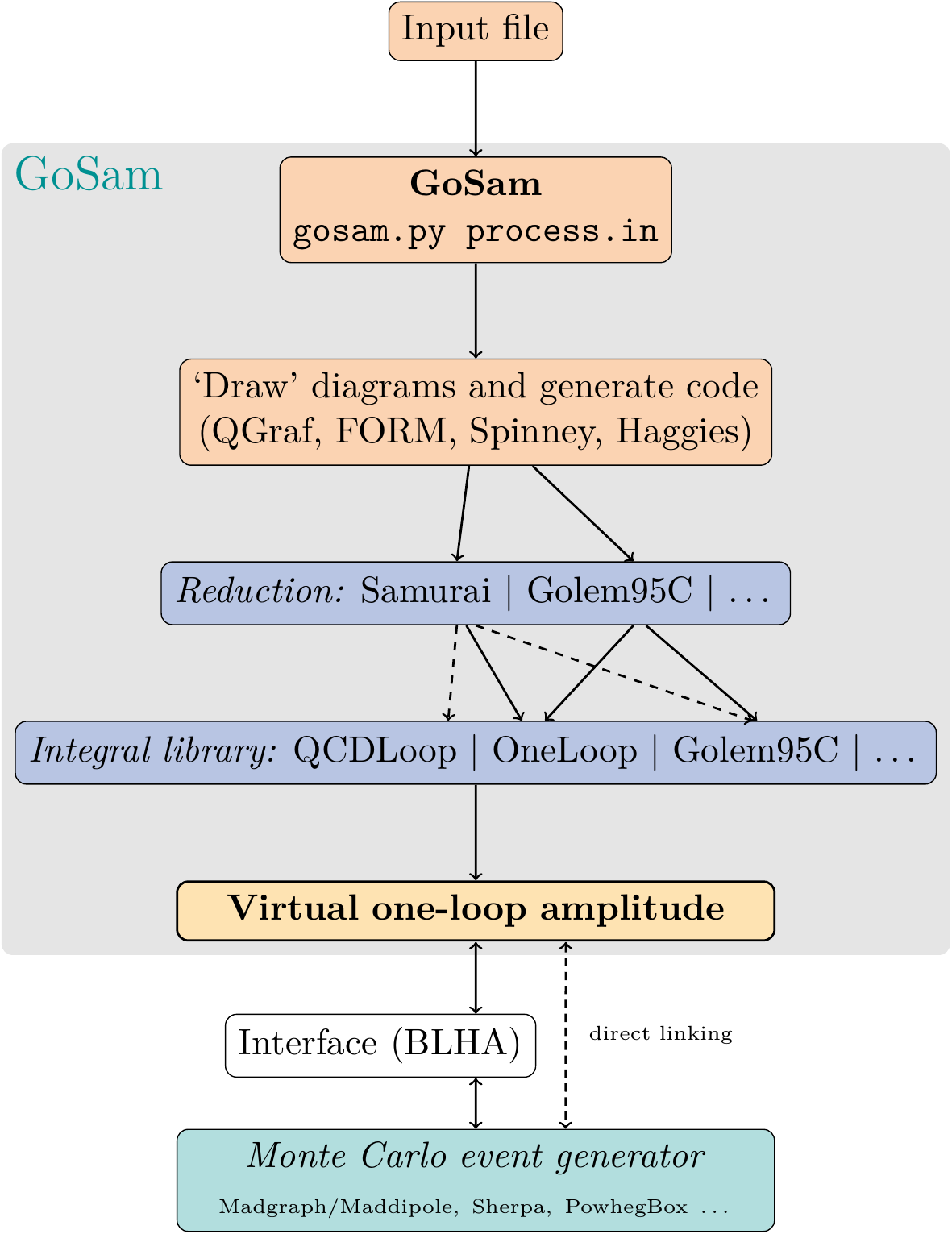}
\caption{Workflow of the program package \GOSAM{}.\label{fig:flowchart}}
\end{figure}

The generated code can then be linked to Monte Carlo programs providing the real emission 
and infrared subtraction parts
and the phase space integration,  either directly  or via
the Binoth Les Houches Interface \cite{Binoth:2010xt}.
A flowchart of the procedure described above is shown in Fig.~\ref{fig:flowchart}.

The tree level, NLO real radiation and the infrared subtraction terms have been produced with 
{\tt MadDipole/MadGraph4}\,\cite{Frederix:2008hu,Frederix:2010cj,Stelzer:1994ta,Maltoni:2002qb,Alwall:2007st}
and have been checked by verifying the independence of the 
result from the unphysical phase space cut parameter  $\alpha$\,\cite{Nagy:1998bb}.
The phase space integration of all ingredients has been performed with {\tt MadEvent}\,\cite{Maltoni:2002qb,Alwall:2007st}.
We have implemented the graviton propagator in the form of eq.~(\ref{eq:propagator})  explicitly 
in the {\tt Helas} routines described in \cite{deAquino:2011ix}.
In order to make sure that the conventions and the implementation of the non-standard propagator 
are consistent between \GOSAM{} and the Monte Carlo program providing the real radiation part, 
the leading order results of \GOSAM{} 
have been compared  with the ones of {\tt MadGraph4}\,\cite{Stelzer:1994ta,Alwall:2007st,deAquino:2011ix} and 
{\tt Sherpa}\,\cite{Gleisberg:2008ta,Gleisberg:2003ue} both at matrix element level and 
at cross section level.

It also has been checked that after UV renormalisation, all poles from the virtual contributions cancel 
with the poles from the infrared insertion operator~\cite{Catani:1996vz} in the real radiation.

\begin{figure}[tbp]
\vspace*{3.cm}
\includegraphics[width=0.9\textwidth,page=1]{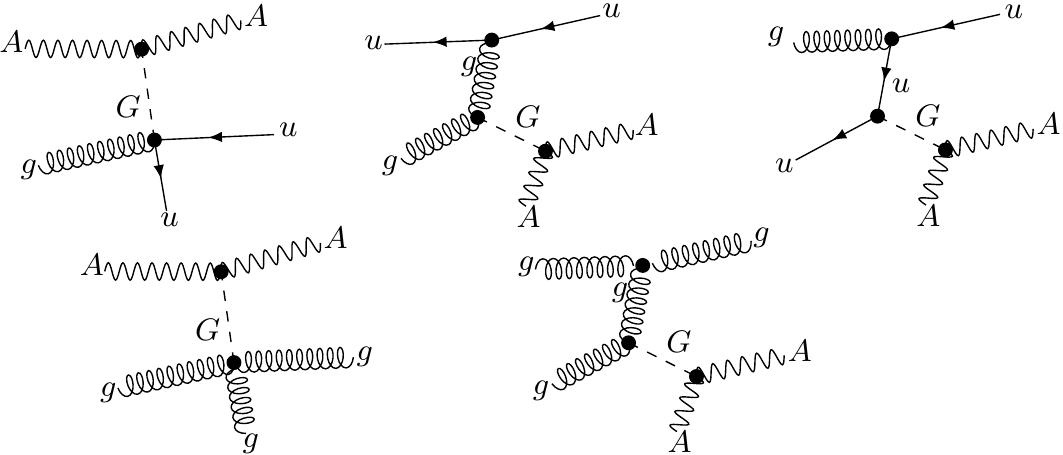}
\caption{Tree level diagrams contributing to the process $pp\to (G \,\to\, \gamma \gamma ) + $jet.
Diagrams which can be obtained by crossing or summing over quark flavours are not shown.\label{fig:LOdiagrams}}
\end{figure}

At tree level, the following subprocesses contribute before summing over flavours
(processes with $p_1\leftrightarrow p_2$ are not shown)
\begin{align*}
	q \bar q \quad &\rightarrow \quad (G \,\to\, \gamma \gamma ) + g \\
	g  q \quad &\rightarrow \quad (G \,\to\, \gamma \gamma ) + q \\
	g  \bar q \quad &\rightarrow \quad (G \,\to\, \gamma \gamma ) + \bar q \\
	g g \quad &\rightarrow \quad (G \,\to\, \gamma \gamma ) + g ,
\end{align*}
where the $gg$ subprocess dominates at the LHC.
The topologically different leading order diagrams are shown in 
Fig.~\ref{fig:LOdiagrams}.

\begin{figure}[tbp]
\vspace*{2cm}
	\includegraphics[width=\textwidth,page=2]{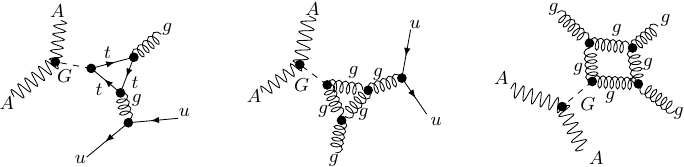}
	\caption{Selected loop level diagrams. The couplings to the spin-two particle (dashed lines)
	lead to tensor ranks exceeding the number of propagators in the loop integrals.\label{fig:loop_level_uux}}
\end{figure}

As the $gq$ and $g  \bar q$ initiated subprocesses can be obtained from the $q\bar q$
initiated one by crossing, \GOSAM{} only generates the virtual corrections
for the $u\bar u$ and $gg$ initial states. The sum over flavours is performed 
when convoluting with the PDFs.
Examples of one-loop diagrams are shown in Fig.~\ref{fig:loop_level_uux}.

The $u\bar u$ initiated subprocess contains 48 diagrams at NLO, the $gg$ initiated one 121~diagrams.
Among the latter are rank five box diagrams, which lead to complicated expressions 
due to the high tensor rank. The program uses the \GOLEMVC{} library, with an extension for 
integrals with tensor ranks exceeding the number of propagators\,\cite{golemhighrank}  
to compute tensor integrals.
Interference between signal and background processes  has been neglected.

The interaction of the gravitons is described by an effective theory which loses its validity 
for partonic energies $\sqrt{\hat{s}}\,\sim M_S$. Therefore we perform the phase space integration 
using a cutoff  $\sqrt{\hat{s}^{\rm{max}}}= M_S-\delta_{\rm{cut}}$ with the default 
value $\delta_{\rm{cut}}=10$\,GeV.
Partial wave analysis in the context of longitudinal $W$ boson 
scattering through the exchange of a spin-2 particle\,\cite{JessicaFrankDiplom} has shown that the unitarity 
bound is not reached below the few TeV range. 
In our case, only couplings of the spin-2 particle to photons and QCD partons enter, therefore we 
do not expect the unitarity bound to set in earlier than in the case considered in Ref\,\cite{JessicaFrankDiplom}.
We refrain from the introduction of form factors to unitarize the amplitude.
Due to the unknown UV completion, the choice of the parameters for the form factors remains ad hoc, except for 
the fact that they should ``smear out" the hard phase space cutoff. 
We checked the cutoff dependence by variation of the cutoff, as shown in Fig.~\ref{fig:cutoff_dep},
and find that the cutoff dependence is rather weak, 
except close to the boundary of phase space $\sqrt{\hat{s}}\,\sim M_S$ where 
the effective theory description is not trustworthy anyway.

%% file: results.tex
We now present phenomenological results for the process 
$pp\to (G\to \gamma\gamma)+$1\,jet+X at NLO, 
where the two photons stem from graviton decay, 
for proton-proton collisions at $\sqrt{s}=8$\Gev. 

\subsection{Setup, input parameters and cuts}

We use the CT10~\cite{Lai:2010vv} parton distributions with 
$N_F=5$ massless flavours and the value of $\alpha_s$ provided by the PDFs. 
For the top mass we take 
$m_t=174\Gev$, the top width is set to zero. 
The jet clustering is done by FastJet \cite{Cacciari:2005hq,Cacciari:2011ma}
using the anti-$k_T$ algorithm~\cite{Cacciari:2008gp} with a cone size of 
$R=0.4$.

The ADD scale $M_S=4\Tev$ and $\delta=4$ extra dimensions are
assumed, unless stated otherwise. 
The renormalisation  and factorisation scales are set dynamically
as
\begin{align}
	\mu_0^2 = \mu_F^2 = \frac14 \left(  m_{\gamma\gamma}^2 +  p_{T,jet}^2  \right)\label{centralscale}
\end{align}
with the invariant mass of the photon pair
\begin{align}
	m_{\gamma\gamma} = \sqrt{ (p_{\gamma_1} + p_{\gamma_2})^2 }\;.
	\label{eq:minvgammagamma}
\end{align}
For the photons, the following cuts are applied:
\be
		p_{T,\gamma} \ge 25\Gev \;,\;
		|\eta_{\gamma}| \le  2.5 \;,\;
		0.4 \le \Delta R_{\gamma\gamma}.
\ee
Additionally, the invariant mass of the photon pair is restricted
to
\begin{align}
	140\Gev \le  m_{\gamma\gamma} < 3.99\Tev.
\end{align}
The lower bound serves to suppress the Standard Model background.
The jets are restricted by the following cuts:
\be 
	 p_{T,\text{leading jet}} \ge 30\Gev  \;,\;
	 |\eta_{\text{jet}}| \le  4 \;,\;
         0.4 \le \Delta R_{\text{jet,$\gamma$}}\;.
\ee
The SM background can be suppressed considerably by 
choosing very large values 
for $p_{T,\gamma}^{\rm{min}}$ and $m_{\gamma\gamma}^{\rm{min}}$. 
In our calculation we choose rather moderate cuts, which are motivated by  the ones used in~\cite{Aad:2012cy}.

\subsection{Results}

\begin{figure}[htb]
\centering
\includegraphics[width=0.7\textwidth]{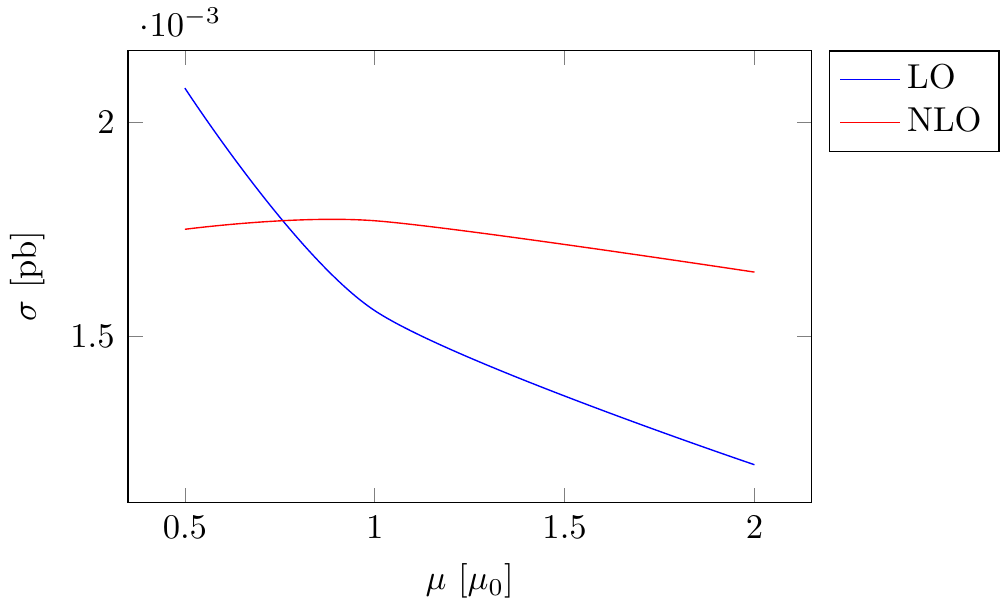}
\caption{Behaviour of the cross section under scale variations, varying by a factor two 
	around the central scale defined in eq.\,(\ref{centralscale}).
	\label{fig:scales}}
\end{figure}

The results for the  total cross sections are shown in Table \ref{tab:totalxs}.
\begin{table}
\centering
\begin{tabular}{r|c|c|cc}
&cross section [fb]&MC error [fb]& \multicolumn{2}{l}{scale uncertainty [fb]}\\
\hline
\hline
LO&1.561&$\pm 6.5\times 10^{-4}$&$^{+0.522}_{-0.363}$&$^{\mu=\mu_0/2}_{\mu=2\mu_0}$\\
\hline
NLO&1.767&$\pm 7.1\times 10^{-3}$&$^{-0.02}_{-0.11}$&$^{\mu=\mu_0/2}_{\mu=2\mu_0}$\\
\end{tabular}
\caption{LO and NLO total cross sections and theoretical uncertainties.}
\label{tab:totalxs}
\end{table}
The scale uncertainty improves significantly between LO and NLO, as can be seen from 
Table \ref{tab:totalxs} and Fig.~\ref{fig:scales}.

One of the most important observables in searches for extra dimensions based on graviton decay into 
two photons is the diphoton invariant mass spectrum, as we expect large enhancements in the tail of the 
$m_{\gamma\gamma}$ distribution.

\begin{figure}[htb]
\centering
\includegraphics[width=0.75\textwidth]{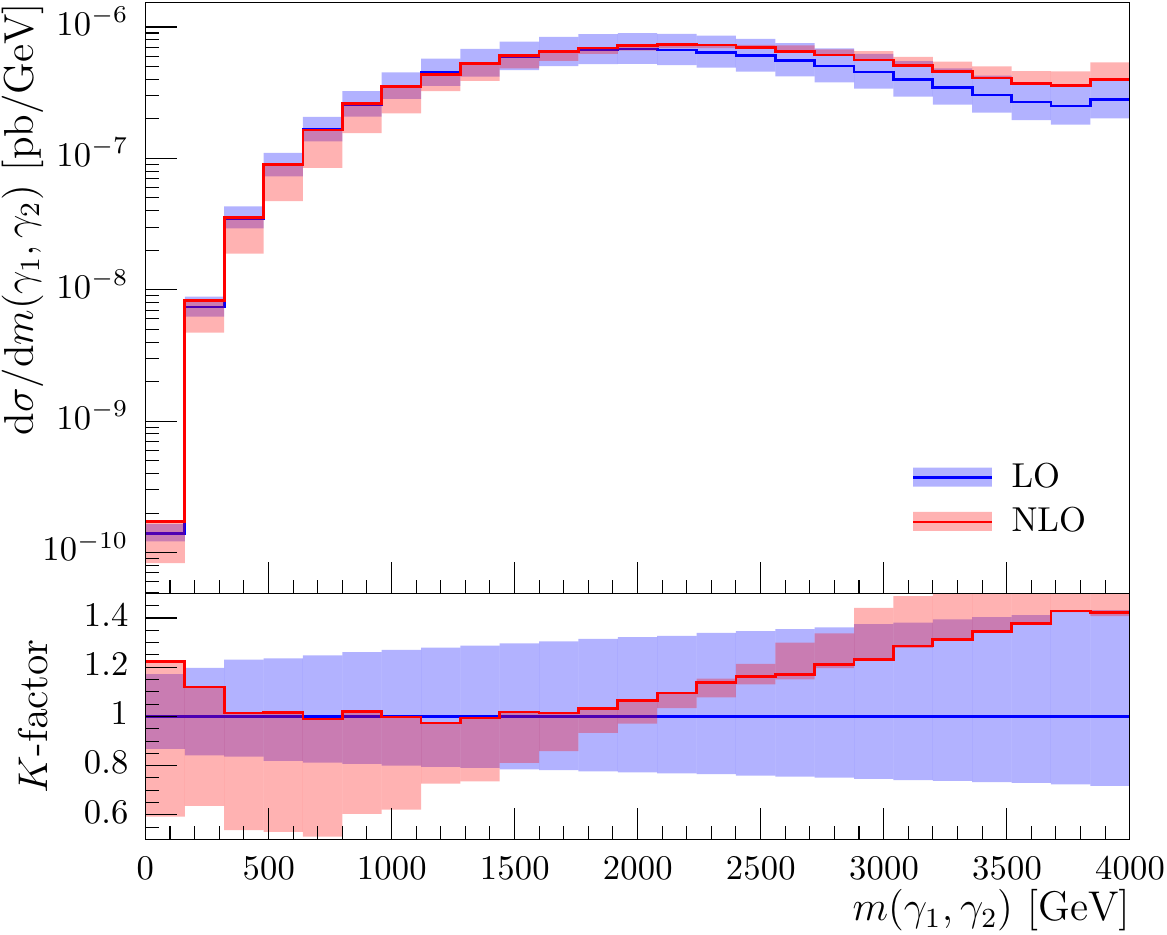}
\caption{NLO QCD corrections to the invariant mass distribution of the photon pair stemming from graviton decay.  
	The bands show the scale variations by a factor of two around the central scale.
	\label{fig:mgg_scalevar}}
\end{figure}

Fig.~\ref{fig:mgg_scalevar} shows the invariant mass distribution of the photon pair at LO and NLO.
We observe that the K-factor is far from being constant for this distribution, increasing 
towards large values of $m_{\gamma\gamma}$.  
In order to verify that this behaviour of the K-factor is not an artifact of the dynamical scale choice, 
we have also made the calculation with a fixed scale of 2~TeV and found a similar behaviour of the K-factor.
Therefore the procedure to take NLO corrections into account by 
rescaling the leading order distribution by a constant K-factor, 
as has been done in all experimental analyses so far, can only give a rough estimate of the NLO corrections.

\begin{figure}[htb]
\centering
\includegraphics[width=0.75\textwidth]{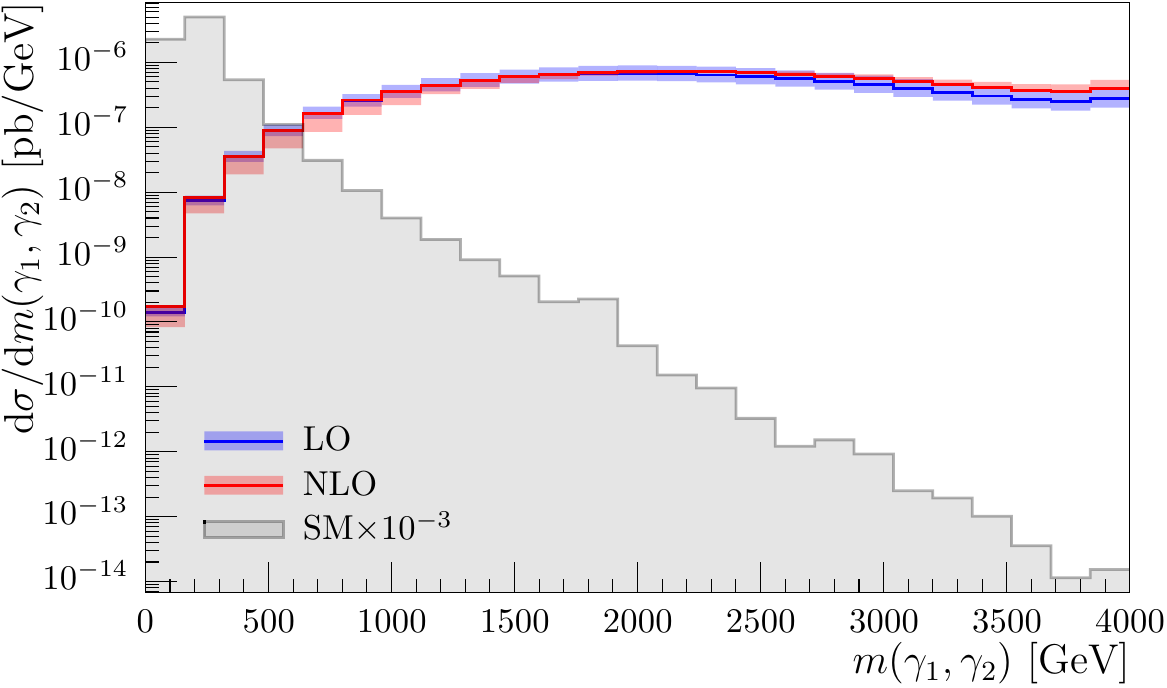}
\caption{Invariant mass of the photon pair at LO and NLO.  
	The bands show the scale variations by a factor of two around the central scale.
	The SM background, rescaled by a factor of $10^{-3}$,  is also shown.
	\label{fig:mgg_plusSM}}
\end{figure}

In Fig.~\ref{fig:mgg_plusSM}  the invariant mass distribution of the photon pair is 
compared to the LO Standard Model background. 
The latter has been rescaled by a factor of $10^{-3}$ in order to be able to compare the shapes.  
It is obvious that the shape of the background for the $m_{\gamma\gamma}$ distribution is very different, 
and an enhancement over the background should be visible in the large $m_{\gamma\gamma}$ region within ADD models. 

\begin{figure}[htb]
\centering
        \begin{subfigure}[]{0.49\textwidth}
                \centering
\includegraphics[width=1\textwidth]{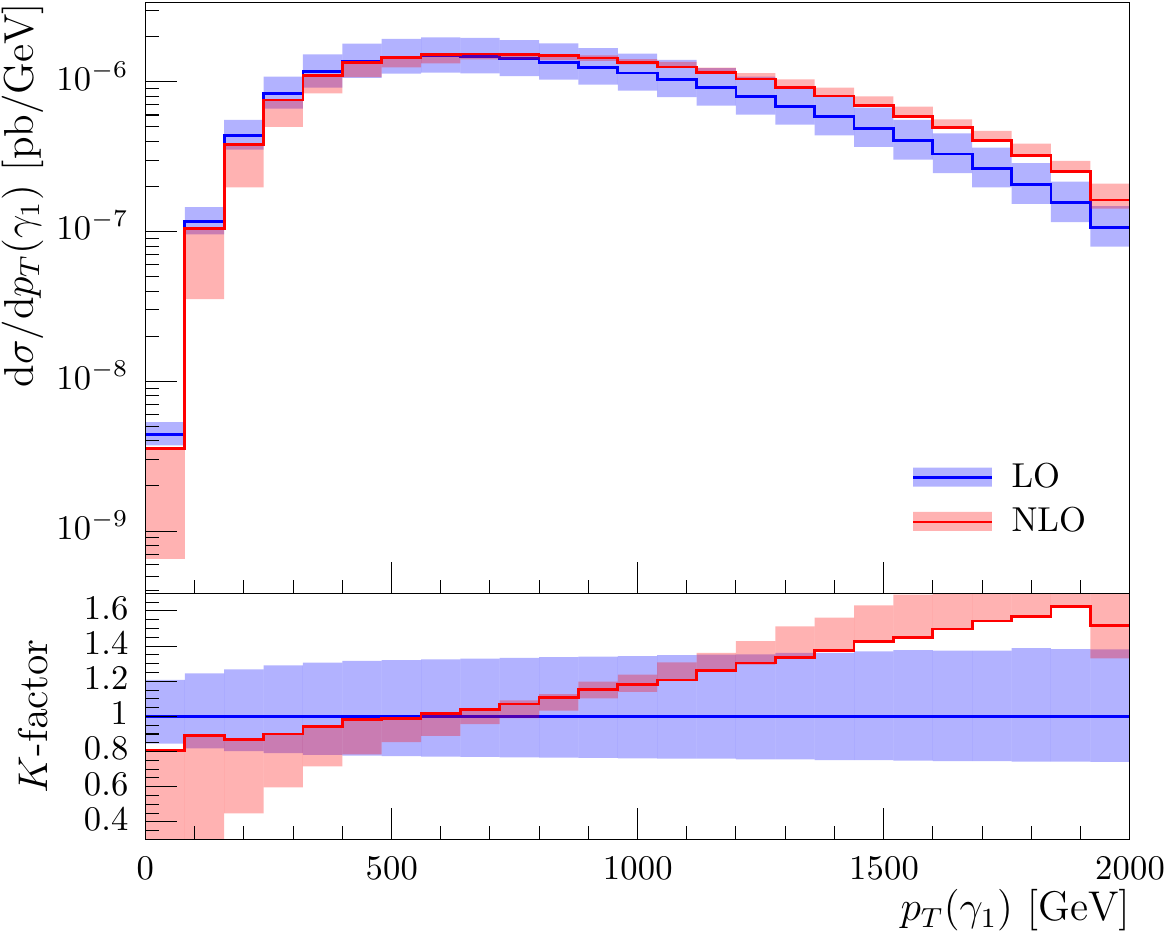}
                \caption{$p_T$ distribution of the leading photon}
                \label{fig:ptgam1_scalevar}
		\end{subfigure}
        \begin{subfigure}[]{0.49\textwidth}
                \centering
\includegraphics[width=1\textwidth]{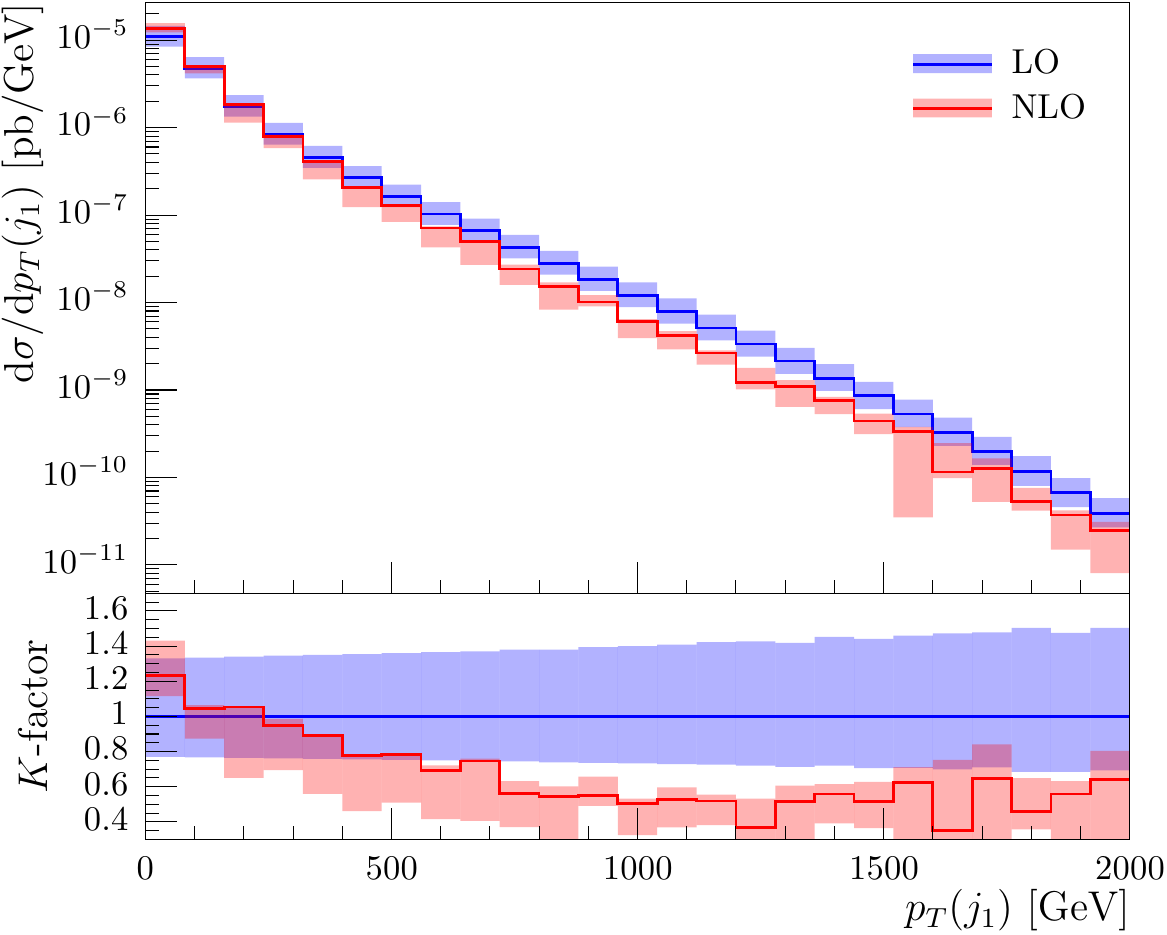}
                \caption{$p_T$ distribution of the jet}
                \label{fig:ptjet_scalevar}
        \end{subfigure}
\caption{NLO QCD corrections to the transverse momentum distribution of (a) the leading-$p_T$ photon 
and (b) the jet.  The bands show the scale variations by a factor of two around the central scale.
\label{fig:ptab}}
\end{figure}

Large variations of the differential K-factor 
can also be observed in the transverse momentum distributions of the photons and of the jet.
The K-factor of the leading-$p_T$ photon, $\gamma_1$, is also increasing towards large $p_T$ values, 
as shown in Fig.~\ref{fig:ptgam1_scalevar}, while the differential K-factor for the 
transverse momentum distribution of the jet is decreasing as the $p_T$ of the jet is increasing, 
see Fig.~\ref{fig:ptjet_scalevar}. The softening of the jet $p_T$ spectrum  at NLO 
can be understood from the fact that at high $p_T$, a single jet (parton) is more likely to radiate 
another parton -- which only is taken into account at NLO -- 
and this makes the original jet softer.

\begin{figure}[htb]
\centering
        \begin{subfigure}[]{0.49\textwidth}
                \centering
\includegraphics[width=1\textwidth]{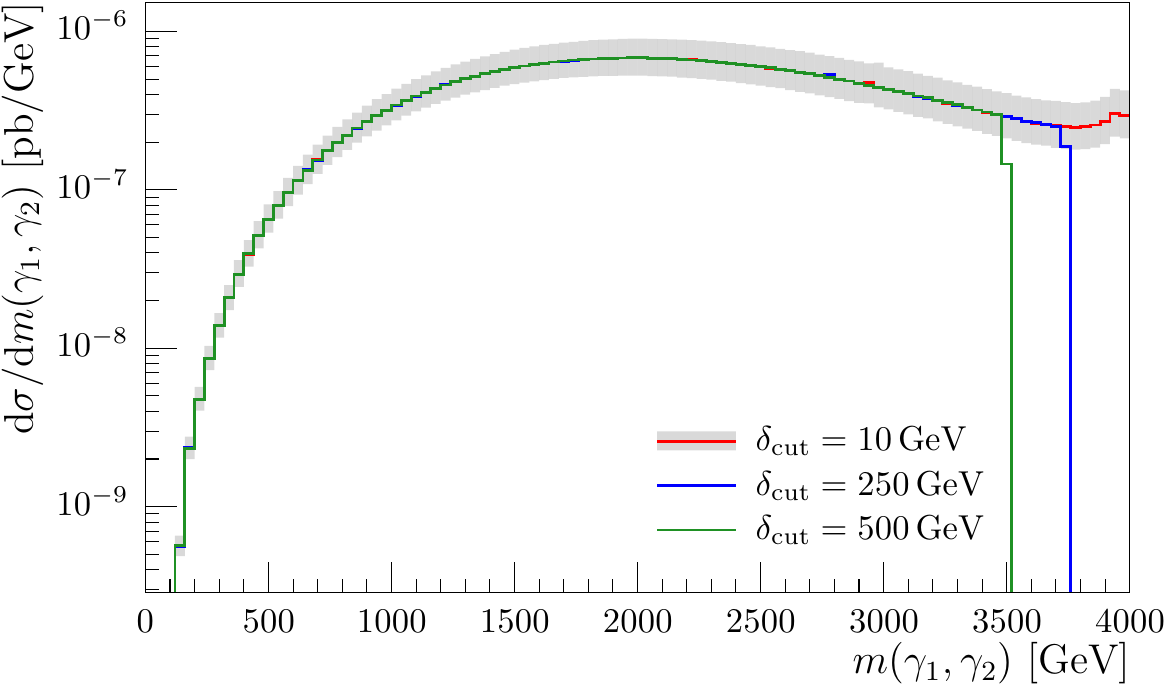}
                \caption{LO}
                \label{fig:cutoffvar_mgg_LO}
        \end{subfigure}
        \begin{subfigure}[]{0.49\textwidth}
                \centering
\includegraphics[width=1\textwidth]{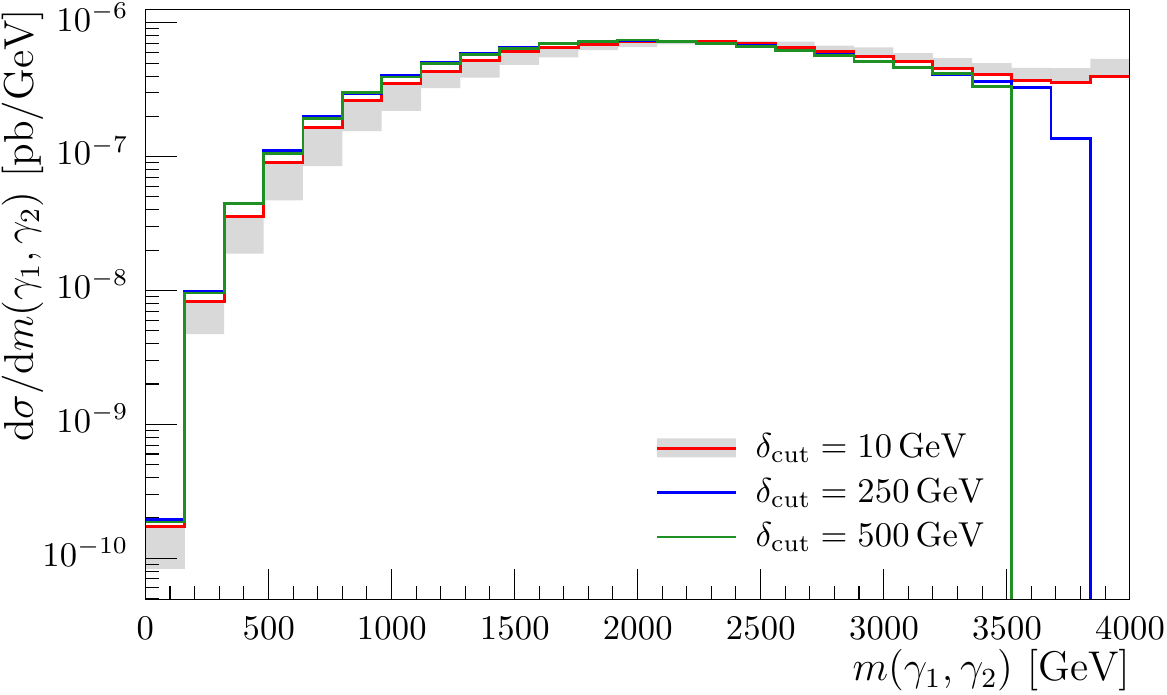}
                \caption{NLO}
                \label{fig:cutoffvar_mgg_NLO}
        \end{subfigure}
\caption{Dependence of the diphoton invariant mass distribution on the value of the cutoff 
		$\sqrt{\hat{s}^{\rm{max}}}= M_S- \delta_{\rm{cut}}$ 
		where the effective theory is expected to loose its range of
		 validity. 
		 \label{fig:cutoff_dep}}
\end{figure}

As we are calculating the QCD corrections to a process involving only graviton 
bridges, while graviton loops are suppressed, the unknown UV completion of the theory 
should not destroy the reliability of the QCD corrections below the scale  $M_S$.
In order to test the dependence on the cutoff scale in the phase space 
integration, we varied 
$\sqrt{\hat{s}^{\rm{max}}}= M_S-\delta_{\rm{cut}}$ using $\delta_{\rm{cut}}=10,250,500$\,GeV.
At LO, except for the region very close to the cutoff, the dependence of the shape on 
$\delta_{\rm{cut}}$ is weaker than the residual scale dependence, 
as can be seen from Fig.~\ref{fig:cutoffvar_mgg_LO}. At NLO, the scale uncertainty is considerably reduced, 
therefore the relative size of the two types of uncertainties does not show this clear hierarchy anymore. 
Nonetheless one can see from Fig.~\ref{fig:cutoffvar_mgg_NLO} that the 
$\delta_{\rm{cut}}$-dependence  is very weak for almost the whole 
range of the distribution, being limited to a small region around the cutoff scale.
Certainly, the size of the total cross section will depend on the cutoff, but 
normalized distributions will show their characteristic shape, independent of the precise value of the cutoff
(as long as it is in the vicinity of $M_S$).
This behaviour confirms that the NLO QCD corrections are not affected to an unacceptable extent by the 
unknown UV completion of the  model. 

\begin{figure}[htb]
\centering
        \begin{subfigure}[]{0.49\textwidth}
                \centering
\includegraphics[width=1\textwidth]{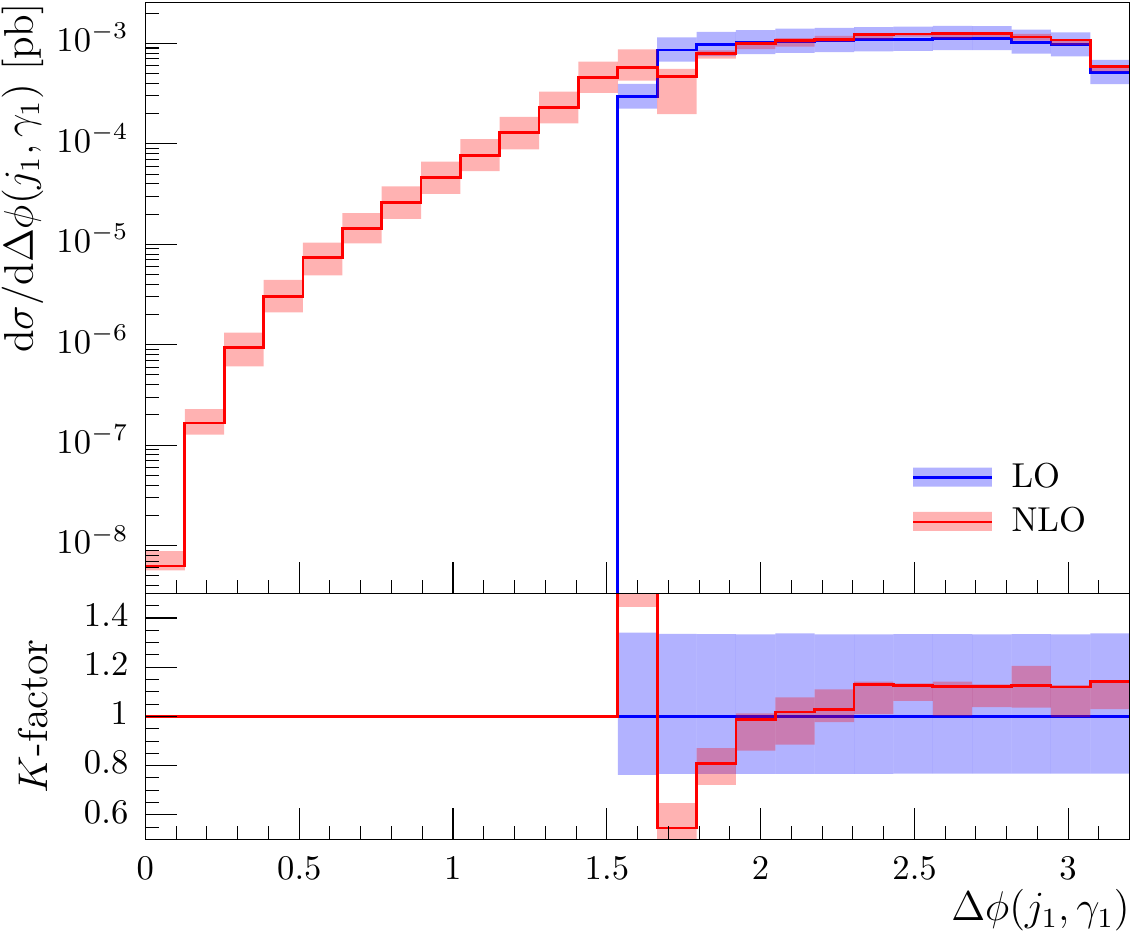}
                \caption{}
                \label{fig:dphijgam1}
        \end{subfigure}
        \begin{subfigure}[]{0.49\textwidth}
                \centering
\includegraphics[width=1\textwidth]{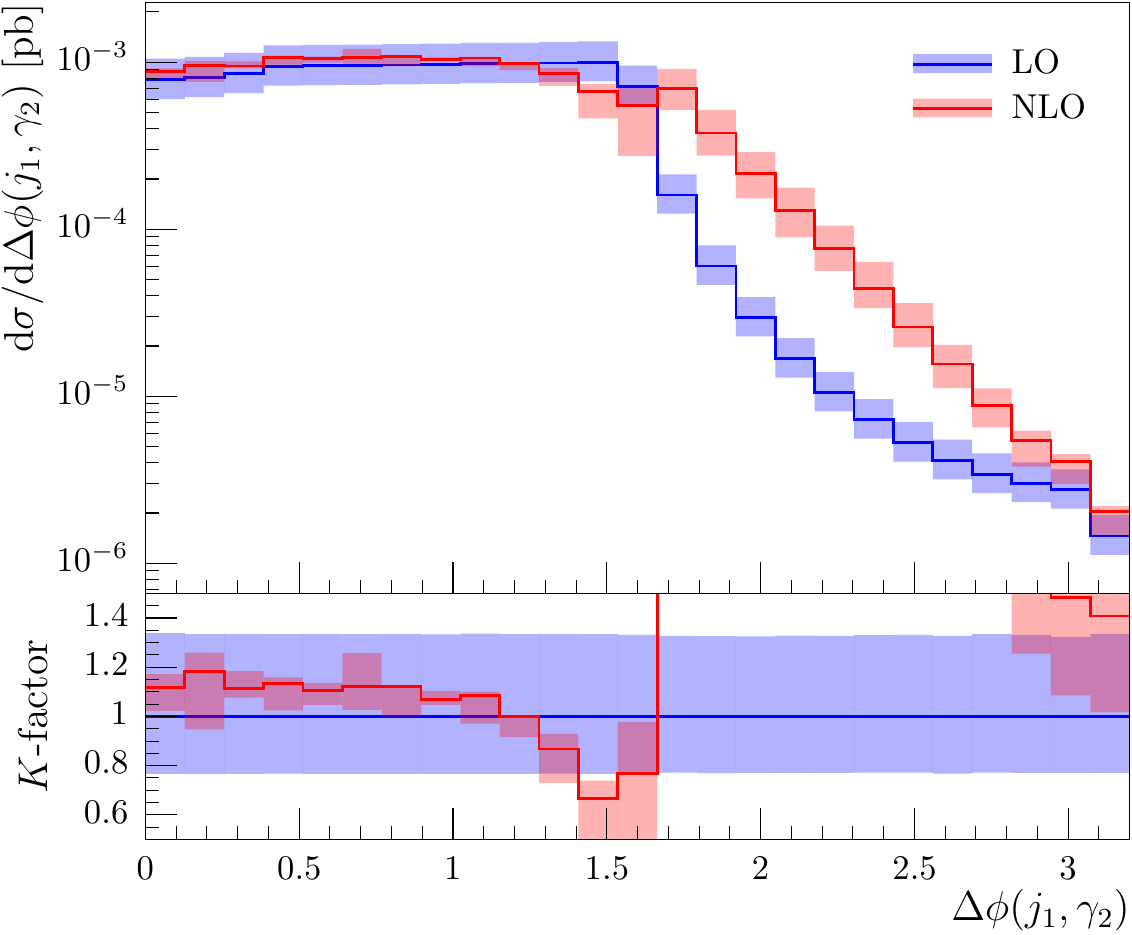}
                \caption{}
                \label{fig:dphijgam2}
        \end{subfigure}
\caption{Distribution of the relative azimuthal angle $\Delta\varphi$ 
between the jet and (a) the leading-$p_T$ photon, (b) the subleading-$p_T$ photon.\label{fig:deltaphi_distrib}}
\end{figure}

Angular distributions are particularly interesting as they can serve to pin down the spin-two nature 
of the object decaying into a photon pair. 
The presence of the jet gives us an extra handle to probe the kinematics. 
The relative azimuthal angle distributions between the jet and the leading respectively subleading photon 
are particularly interesting. 
Fig.~\ref{fig:deltaphi_distrib} illustrates 
that the NLO corrections significantly 
alter the shape because the extra parton present in the real radiation 
contribution opens up a region which is kinematically inaccessible at LO. 
Even though this feature certainly is also present in the SM background, 
Fig.~\ref{fig:deltaphi_distrib} clearly displays the importance of NLO corrections.

\begin{figure}[htb]
\centering
        \begin{subfigure}[]{0.49\textwidth}
                \centering
\includegraphics[width=1\textwidth]{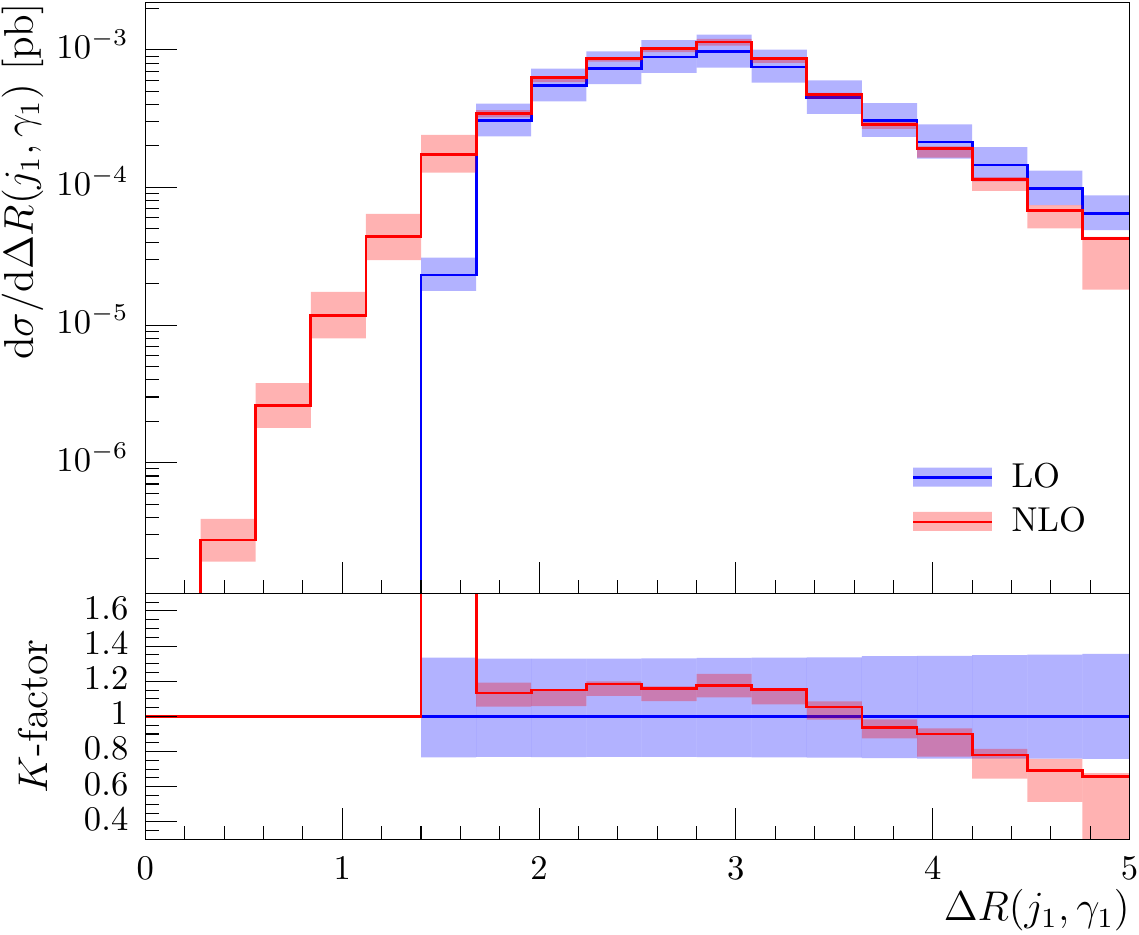}
                \caption{}
                \label{fig:dRjgam1}
        \end{subfigure}
        \begin{subfigure}[]{0.49\textwidth}
                \centering
\includegraphics[width=1\textwidth]{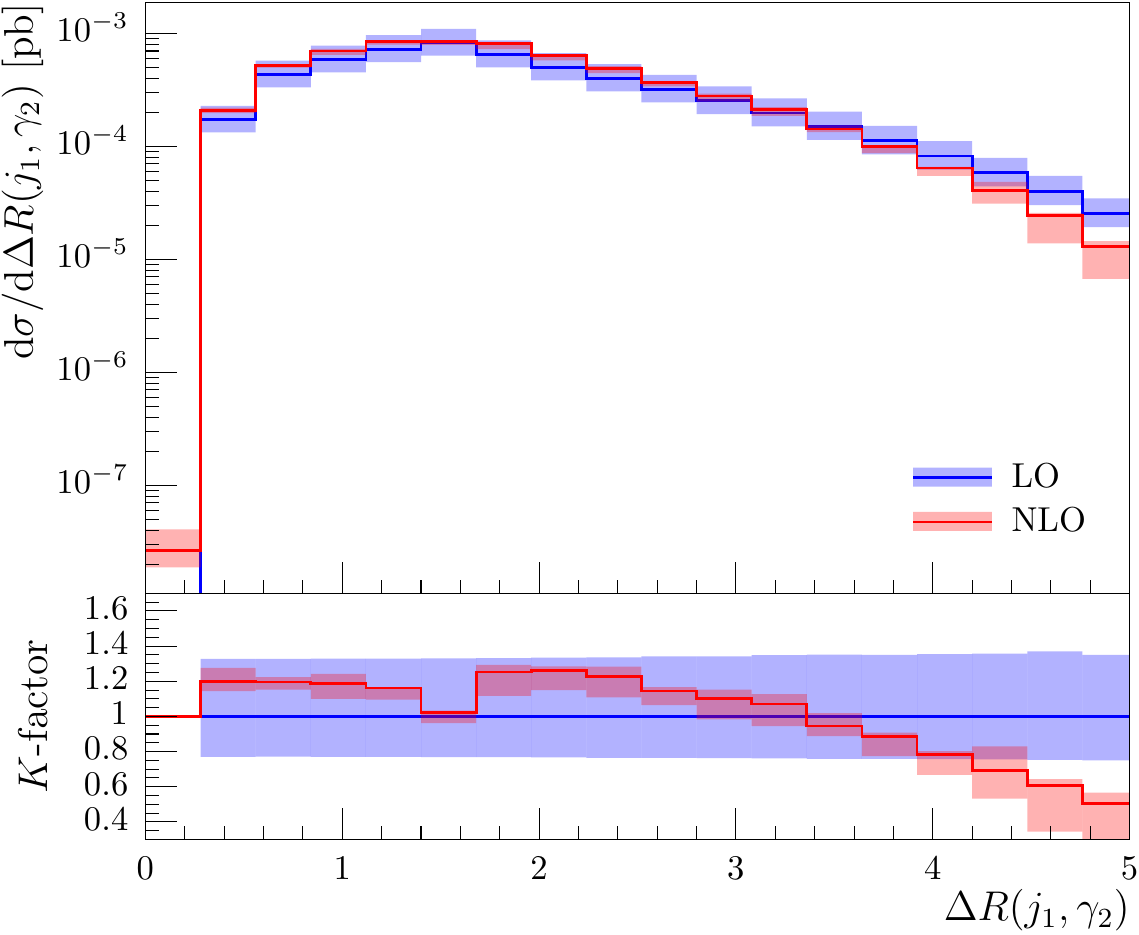}
                \caption{}
                \label{fig:dRjgam2}
        \end{subfigure}
\caption{Distribution of the distance  $\Delta R$ 
between the jet and (a) the leading-$p_T$ photon, (b) the subleading-$p_T$ photon.\label{fig:deltaR_distrib}}
\end{figure}

\begin{figure}[htb]
\centering
        \begin{subfigure}[]{0.49\textwidth}
                \centering
\includegraphics[width=1\textwidth]{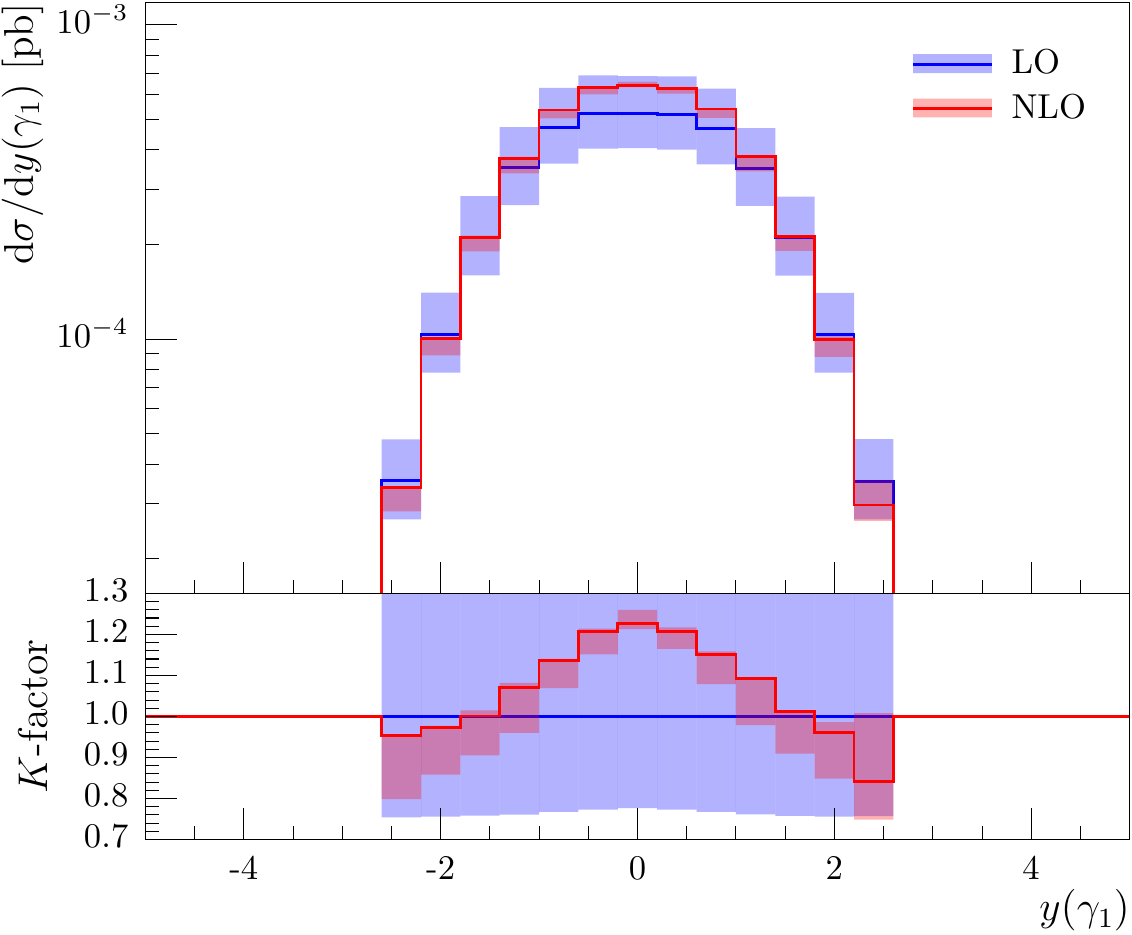}
                \caption{}
                \label{fig:dygam1}
        \end{subfigure}
        \begin{subfigure}[]{0.49\textwidth}
                \centering
\includegraphics[width=1\textwidth]{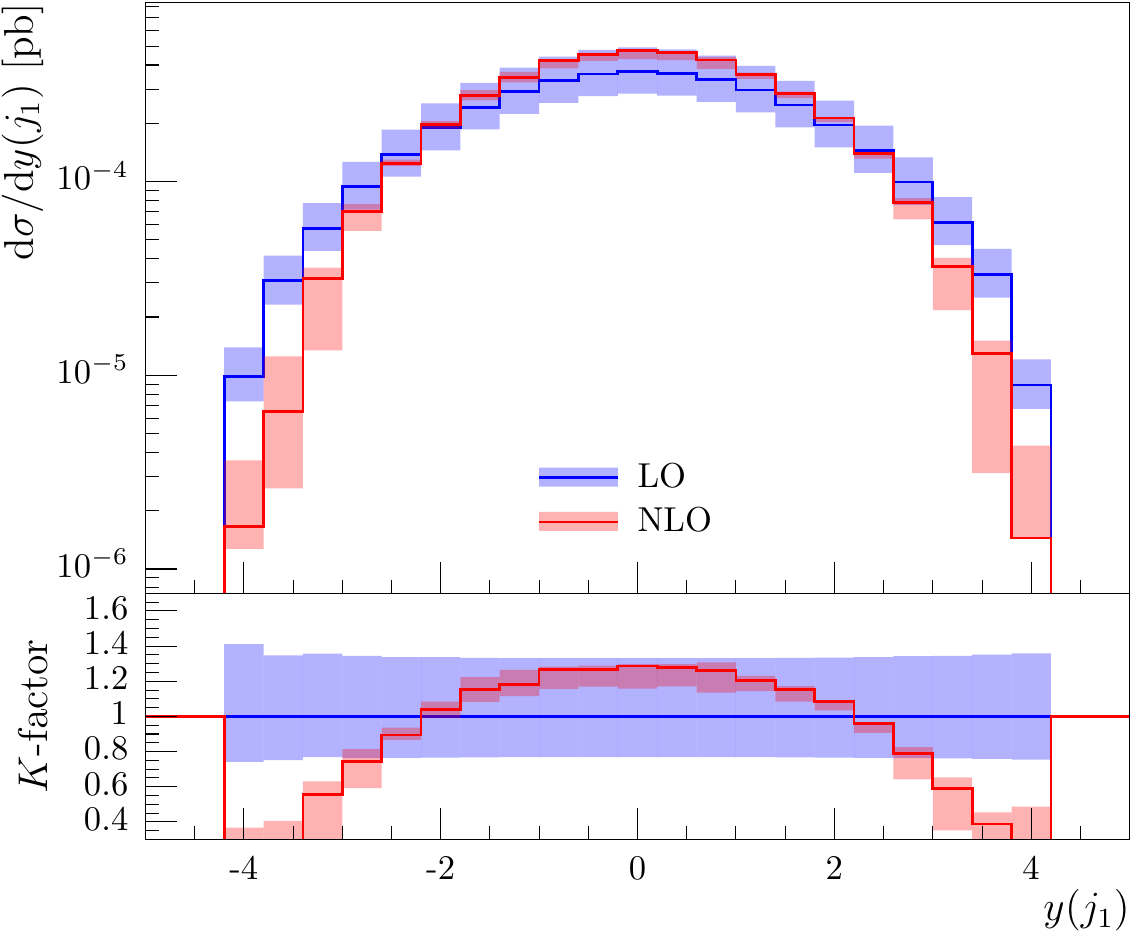}
                \caption{}
                \label{fig:dyjet}
        \end{subfigure}
\caption{Rapidity distributions for (a) the leading-$p_T$ photon, (b) the jet.\label{fig:rap_distrib}}
\end{figure}

In Fig.~\ref{fig:deltaR_distrib} the distributions for the distance in rapidity and azimuthal angle space 
$\Delta R(j,\gamma)=\sqrt{(\eta^j-\eta^\gamma)^2+(\varphi^j-\varphi^\gamma)^2}$ between 
the jet and the leading and subleading-$p_T$ photon are shown.
Fig.~\ref{fig:rap_distrib} displays the rapidity distributions of the leading photon respectively the jet.
Again, the non-uniform K-factors are clearly visible.

\begin{figure}[htb]
\centering
        \begin{subfigure}[]{0.49\textwidth}
                \centering
\includegraphics[width=1.\textwidth]{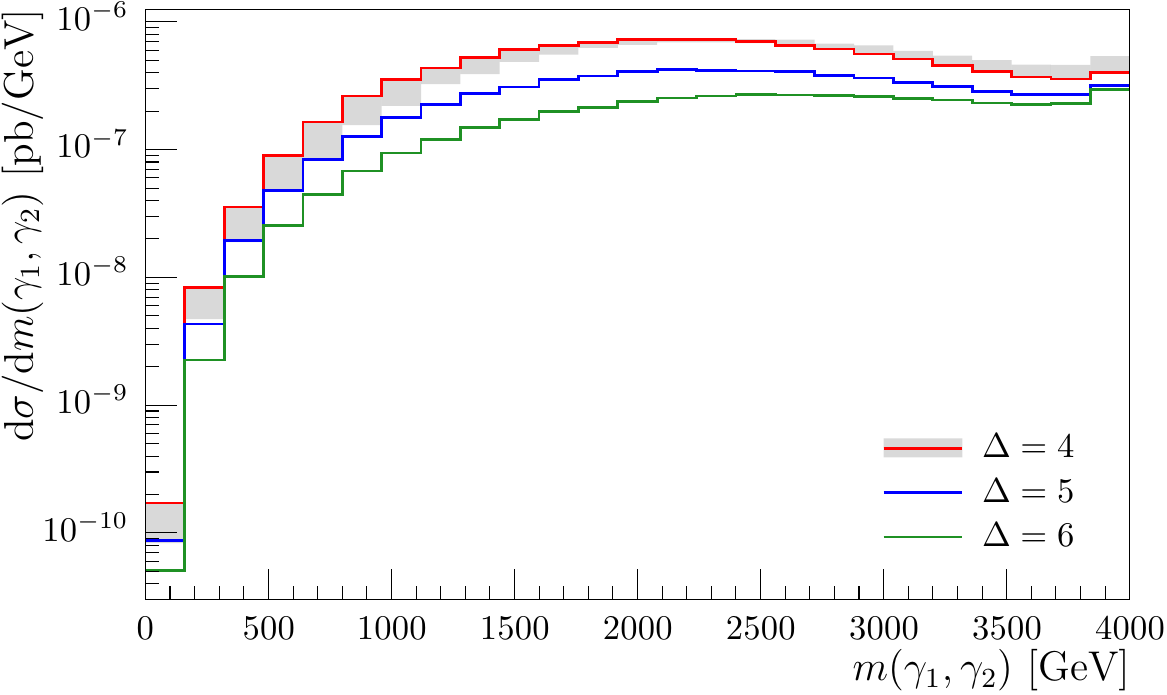}
                \caption{}
                \label{fig:deltavarmgg}
        \end{subfigure}
        \begin{subfigure}[]{0.49\textwidth}
                \centering
\includegraphics[width=1.\textwidth]{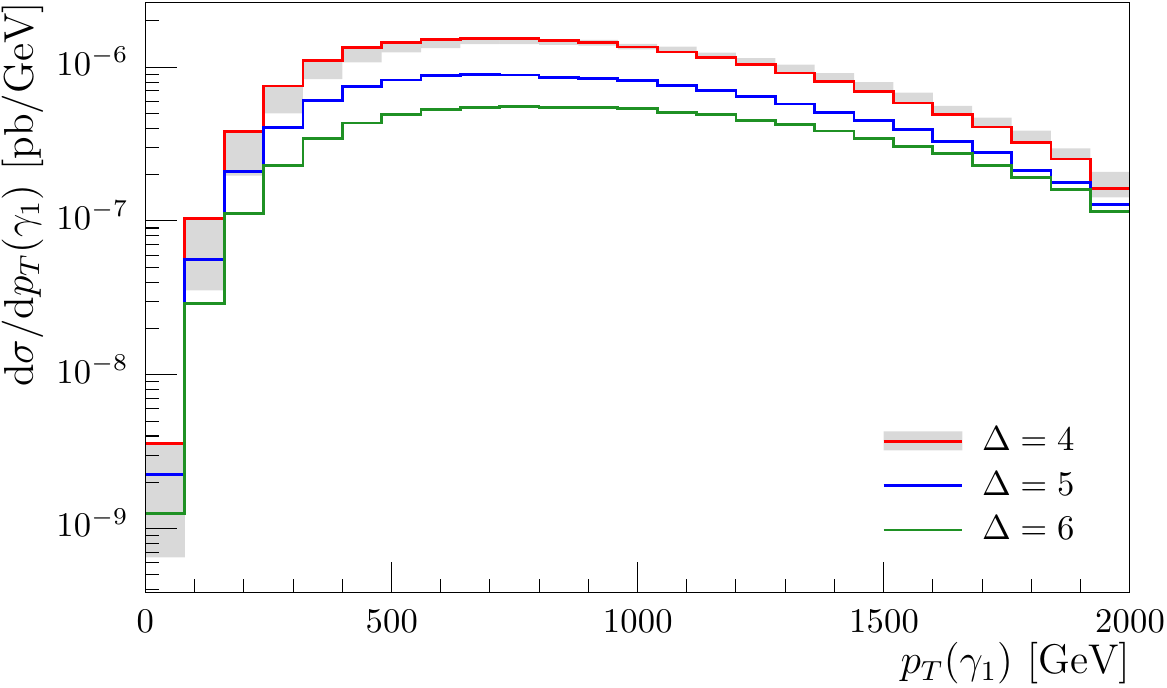}
                \caption{}
                \label{fig:deltavarptgam1}
        \end{subfigure}
\caption{(a) Diphoton invariant mass and (b) transverse momentum distribution 
for different numbers of extra dimensions at NLO. The scale uncertainty band is only included for 
$\Delta=4$.
\label{fig:deltavar1}}
\end{figure}

\begin{figure}[htb]
\centering
        \begin{subfigure}[]{0.49\textwidth}
                \centering
\includegraphics[width=1.\textwidth]{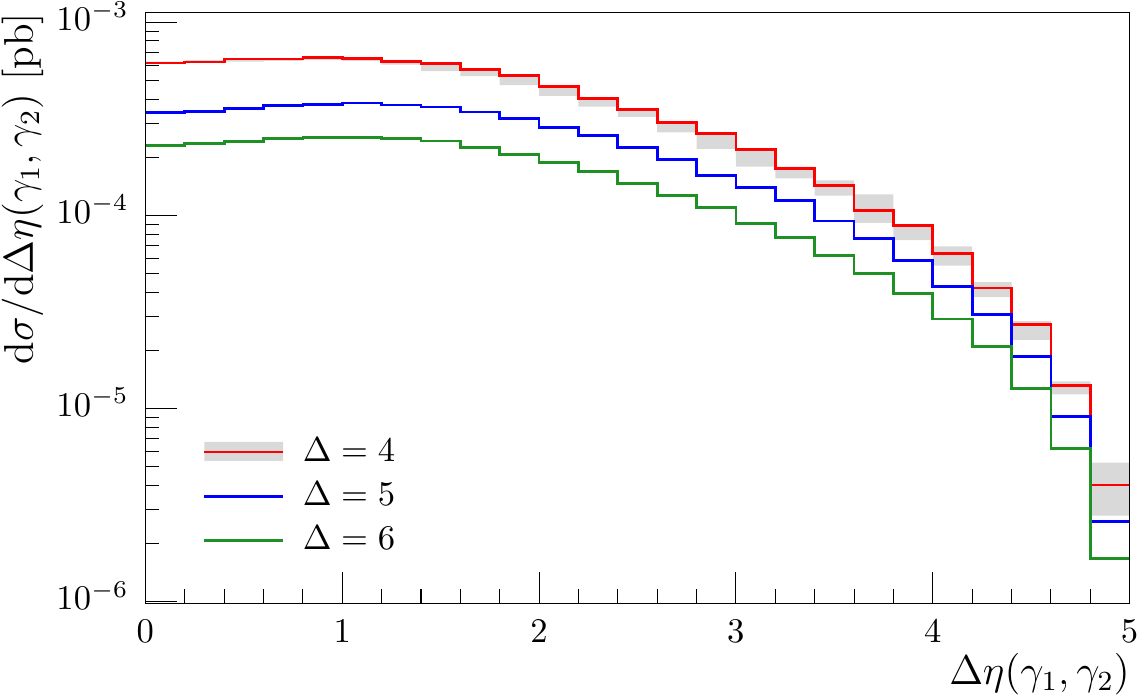}
                \caption{}
                \label{fig:deltavar_eta}
        \end{subfigure}
        \begin{subfigure}[]{0.49\textwidth}
                \centering
\includegraphics[width=1.\textwidth]{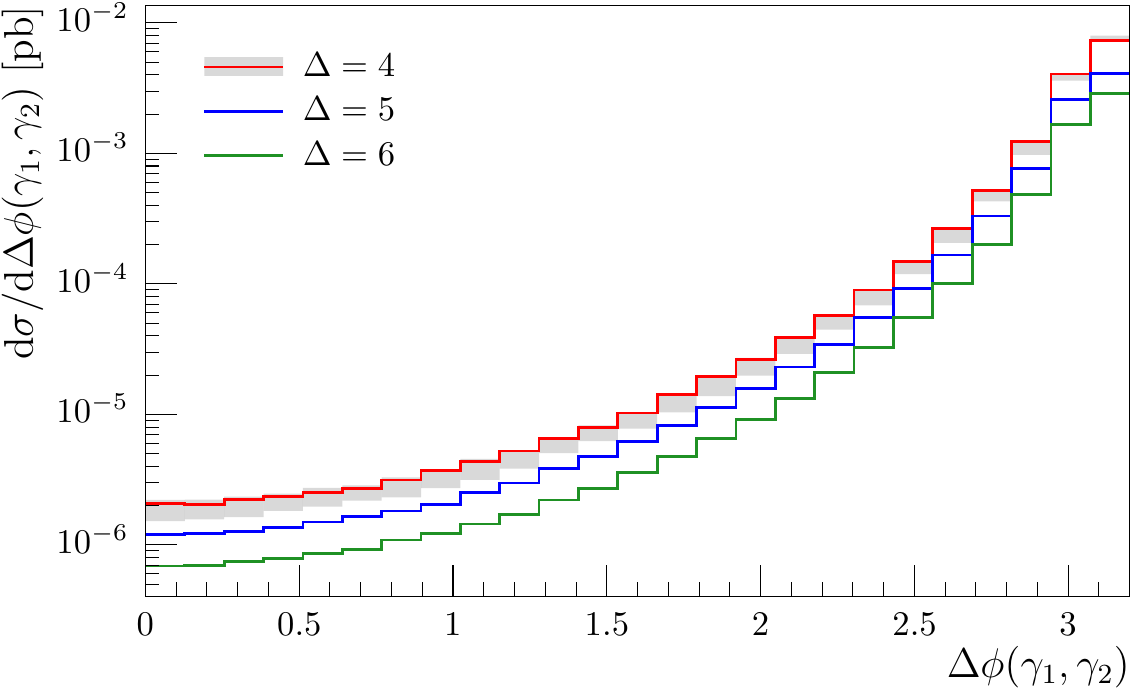}
                \caption{}
                \label{fig:deltavar_}
        \end{subfigure}
\caption{(a) Diphoton rapidity difference and (b) relative azimuthal angle distributions 
for different numbers of extra dimensions at NLO.\label{fig:deltavar2}}
\end{figure}

Finally, we also investigate the case of five or six extra dimensions rather than four.
The higher the number of extra dimensions, the weaker the exclusion limits, 
as the total cross section decreases with increasing dimensions. 
As can be seen from Figs.~\ref{fig:deltavar1} and \ref{fig:deltavar2}, the qualitative 
behaviour is rather similar  for different numbers of extra dimensions, 
even though the propagator has a different analytic form 
for odd numbers of extra dimensions.

%% file: conclusion.tex
We have calculated the NLO QCD corrections to the process 
$pp \to (G\to \gamma\gamma)+$\,jet+X within the ADD~\cite{ArkaniHamed:1998rs}
model of large extra dimensions. 

The one-loop part of the calculation has been provided by the automated program package \GOSAM{}, 
which is publicly available at {\tt http://gosam.hepforge.org/}.
\GOSAM{} can import any Beyond the Standard Model file in 
 \UFO \,(Universal Feynrules Output) format and can deal with 
 effective vertices and particles up to spin two. 
For the real radiation parts we used the
{\tt MadDipole/MadGraph4} framework.
As the effective field theory approximation breaks down for energies 
exceeding the fundamental scale of quantum gravity 
(which is supposed to be in the TeV range for ADD models), 
the phase space integrations have been restricted to center of mass energies 
below the scale $M_S\sim 4$\,TeV. We have studied the UV cutoff dependence in detail and 
demonstrated that the latter does not affect the distributions significantly, except for a 
small region close to the cutoff value.

The corrections significantly reduce the 
scale uncertainty and lead to K-factors for the total cross section of the order of 1.1 to 1.4. 
It is important to note that the differential K-factors are far from being constant. 
For the diphoton invariant mass distribution as well as for the photon transverse momentum
distributions the K-factors 
increase  up to about 1.5 towards the tail of the distributions, 
while  for the jet $p_T$ distribution, the behaviour is the opposite.

We have studied the cases $\delta=4,5,6$ extra dimensions and find similar qualitative behaviour, 
while the cross sections are decreasing as the number of extra dimensions grows.
We also investigate angular distributions where the NLO corrections significantly change the shape,
as kinematic regions open up which are not accessible at leading order.

This calculation illustrates the power and flexibility of \GOSAM{} to do one-loop 
calculations for multi-particle final states stemming from interactions beyond the Standard Model. 
Further applications in this direction will hopefully be confronted with data hinting to BSM physics
in the not too distant future.

%% file: gravi_main.bbl
\providecommand{\href}[2]{#2}\begingroup\raggedright\endgroup